\documentstyle[aps,manuscript]{revtex}
\begin{document}
\title{Variational study of  dilute Bose condensate in a harmonic trap}
\author{Alexander L. Fetter}
\address{Departments of Physics and Applied Physics, Stanford University,
Stanford, CA 94305-4060}
\date{\today}
\maketitle
\begin{abstract}
 A two-parameter trial condensate wave function is used to find an
approximate variational solution to the
Gross-Pitaevskii equation for $N_0$ condensed bosons in an isotropic
harmonic trap with oscillator length $d_0$  and
interacting through a repulsive two-body scattering length $a>0$.  The
dimensionless parameter ${\cal N}_0 \equiv
N_0a/d_0$ characterizes the effect of the interparticle interactions, with
${\cal N}_0 \ll 1$ for an ideal gas and ${\cal
N}_0 \gg 1$ for a strongly interacting system (the Thomas-Fermi limit).
The trial function
interpolates smoothly between these two limits, and the  three separate
contributions (kinetic
energy, trap potential energy, and two-body interaction energy) to the
variational condensate energy and the condensate
chemical potential are
 determined parametrically for any value of
${\cal N}_0$, along with illustrative numerical values. The straightforward
generalization to an anisotropic harmonic
trap is considered briefly.

\end{abstract}
\section{introduction}
The remarkable observation of Bose-Einstein condensation of dilute
ultracold alkali atoms in a harmonic trap
\cite{And,Brad,Dav}
 has stimulated much theoretical activity,  based mostly on the
Gross-Pitaevskii (GP)  equation \cite{EPG,LPP} that is
believed to describe accurately the experimental situation well below the
onset of Bose-Einstein condensation. In
this picture, the two-body interaction is characterized by the \(s\)-wave
scattering length \(a\) (typically,
\(a \sim 10~\rm nm\) and positive, but \(a\) is negative for \(^7\)Li \cite
{Brad}). The actual traps are anisotropic,
but, for simplicity, the present work concentrates on  an isotropic trap
with frequency \(\omega_0\) and
oscillator length
$d_0=\sqrt{\hbar/m\omega_0}$ (typically, $d_0 \sim {\rm~a~few~\mu m}$).
Although  the nonlinear GP  equation is readily
solved numerically for the selfconsistent wave function
\(\Psi\) of
\(N_0 \)  condensed atoms  in an isotropic trap \cite{EB,RHBE} with a total
of \(N\) atoms, it is often useful to have
an  analytic approximation  that provides a semiquantitative description of
the condensate density \(n_0({\bf r})\)
associated with the \(N_0\le N\) condensed atoms.  For example, Baym and Pethick
\cite{BP} proposed two very different approximations:  a Gaussian trial
function
 in the limit of a small condensate (\({\cal N}_0 \equiv N_0|a|/d_0\ll 1\))
and a  Thomas-Fermi (TF) approximation in the
limit of a large condensate (\({\cal N}_0 \equiv N_0|a|/d_0\gg 1\)).  The
present work studies a simple variational trial
function that includes  the  two limiting cases and yields a useful
interpolation for intermediate values of
\({\cal N}_0 \).

The basic formalism is reviewed in Sec.~II, and the explicit variational
trial function is analyzed in Sec.~III for
positive scattering length \(a > 0\).  Typical numerical values are
presented in Sec.~IV, along with the
  various components of the energy per particle and the chemical potential.
The quite different situation for negative
scattering length is discussed briefly in Sec.~V, and the generalization to
an anisotropic harmonic trap is considered
in Sec.\ VI.
\section{basic formalism}
In the special case of a uniform dilute Bose gas at $T = 0$ K subject to
periodic boundary
conditions, Bogoliubov \cite{Bog} made the essential  observation  that the
total number of condensed atoms is
large (\(N_0~\gg~1\)).  More generally, for a nonuniform condensate, the
occupation of the various single-particle states
can be characterized by creation and annihilation operators
\(a_j^{\dagger}\) and \(a_j\) that obey Bose commutation
relations \([a_j,a_k^{\dagger}] = \delta_{jk}\).  The condition
\(N_0\gg 1\) implies that the operators for the condensate mode can be
treated simply as numbers \cite{Bog,FW}, because
their commutator
\([a_0,a_0^{\dagger}]= 1\) is small compared to their value \(a_0 \approx
a_0^{\dagger}\approx \sqrt{N_0}\).  As a result, the condensate is
described by a conventional  wave function
\(\Psi{(\bf r})\) and a  ``thermodynamic-potential''  functional [obtained
as a Legendre transformation from the
``energy'' functional \(E_0(N_0)\)]
\begin{equation}K_0(\mu_0) \equiv  E_0-\mu_0N_0=\int
d^3r\,\Psi^*\,(T+U-\mu_0)\,\Psi+2\pi a\hbar^2m^{-1}\int
d^3r\,|\Psi|^4 \label{eq:func}\end{equation}
that gives the dominant contribution for small noncondensate number
\(N'\equiv N - N_0\ll N\).   Here \(E_0(N_0)\) is the
condensate energy,
\(T = -\hbar^2\nabla^2/2m\) is the kinetic energy and
\(U({\bf r}) =
\frac{1}{2}m\omega_0^2r^2\) is the isotropic harmonic trap potential.  The
condensate wave function is normalized to the
total condensate number
\begin{equation}\int d^3r\,|\Psi|^2 = N_0,\label{eq:norm}\end{equation}
so that the condensate particle density is simply \(n_0({\bf r}) =
|\Psi({\bf r})|^2\);
the Euler-Lagrange equation for the functional in  Eq.~(\ref{eq:func})  is
just   the nonlinear Gross-Pitaevskii (GP)
eigenvalue  equation
\begin{equation}(T + U +V)\Psi= \mu_0\Psi,\label{eq:GP}\end{equation}
where \(V({\bf r})  = 4\pi a \hbar^2n_0({\bf r})/m\) is the mean
(pseudo)potential energy per condensed particle. In the
present case of a harmonic trap,  \(\mu_0\)
must be adjusted to ensure that the wave function obeys the asymptotic
boundary condition that \(\Psi \to 0\) as \(r \to
\infty\).  In addition, \(\mu_0\)  obeys the thermodynamic relation
\begin{equation}\mu_0 = \frac{\partial E_0(N_0)}{\partial
N_0},\label{eq:mu}\end{equation}
or, equivalently,
\begin{equation}N_0 = -\frac{\partial K_0(\mu_0)}{\partial \mu_0}.\end{equation}
If the exact condensate wave function is known, then the exact condensate
energy is given by
\begin{equation}E_0=\langle T\rangle + \langle U\rangle
+{\textstyle\frac{1}{2}}\langle
V\rangle,\label{eq:en}\end{equation}
where the angular brackets denote an expectation value evaluated with the
condensate wave function
(namely, \(\langle\cdots\rangle = \int d^3r\,\Psi^*\cdots\Psi\)).  A
combination of Eqs.~(\ref{eq:mu})
and (\ref{eq:en}) gives
\begin{equation}\mu_0N_0 = \langle T\rangle + \langle U\rangle +\langle
V\rangle,\label{eq:mu0}\end{equation}
because \(V\) itself is proportional to \(N_0\) [this last result also
follows directly from the GP Eq.~(\ref{eq:GP})].
\section{variational trial function}
It is convenient to introduce dimensionless units, with \(d_0\)  and
\(\hbar\omega_0\) as the length and energy scales.
In this case, the dimensionless kinetic energy and trap potential energy
become \(T = -\frac{1}{2}\nabla^2\) and \(U =
\frac{1}{2}r^2\).  As a simple  variational trial function for the
isotropic condensate, take
\begin{equation}\Psi(r)=c_0\,(1-r^2/R^2)^{(1+\lambda)/2}\,\theta(R - r)
\label{eq:psi}\end{equation}
that depends on the two dimensionless parameters \(\lambda\) and \(R\).
The normalization constant \(c_0\)  follows immediately from
Eq.~(\ref{eq:norm})
\begin{equation}c_0^2 = \frac{N_0\,\Gamma(\frac{7}{2} + \lambda)}{2\pi
R^3\,\Gamma(\frac{3}{2})\,\Gamma(2+
\lambda)}.\label{eq:c0}\end{equation}

For \(\lambda\to 0\), the trial function obviously reduces to the TF
approximation
\(\Psi_{TF}(r)\propto\big(1-r^2/R^2\big)^{1/2}\,\theta(R-r)\), which
follows by neglecting the kinetic energy
relative to the trap potential energy and the interparticle potential
energy;  in this case  \(U -\mu_0 + V \)
vanishes in the region where $\Psi \neq 0$, and
\(\mu_0\approx
\frac{1}{2} R^2\) for an isotropic harmonic trap.   The condensate density
in this TF  limit is parabolic, with \(n_0(r)
\propto (1-r^2/R^2)\,\theta(R-r)\), showing that $R$ is the (large)
dimensionless TF condensate radius, and  it is easy to
verify that
\cite{BP}  \(R^5 \approx 15{\cal N}_0\),
where \({\cal N}_0 \equiv N_0a/d_0\gg 1\), and
that \(E_0/N_0 \approx \frac{5}{7} \mu_0\).

The opposite case occurs for \(\lambda\to \infty\).  In this limit, a
detailed analysis (see below) shows that
\({\cal N}_0 \approx \sqrt{32\pi}/\lambda\ll 1\), and that \(R\approx
\sqrt{\lambda}\gg 1\);  in addition, a standard
asymptotic  identity involving the Gamma function \cite{AS}
\begin{equation}\frac{\Gamma(z + a)}{\Gamma(z+b)} \sim z^{a-b}\,\big[1 + {\cal
O}(z^{-1})\big],\quad\hbox{for \(|z| \to
\infty\)},\label{eq:Gamma}\end{equation}
verifies that the trial function reduces to a normalized
isotropic Gaussian with condensate density \(n_0(r) = |\Psi_G(r)|^2 =
N_0\pi^{-3/2}\,\lim_{\lambda\to\infty}\,\big(1-r^2/\lambda\big)^{1+\lambda}
\approx N_0\pi^{-3/2}\,\exp(-r^2)\), as used by
Baym and Pethick \cite{BP} (in this low-density, near-ideal,  limit, the
condensate density is  independent of
$R$, and  the mean-square condensate radius is  $\frac{3}{2}$).

For general \(\lambda\) and \(R\), the various expectation values are
readily evaluated:
\begin{equation}\langle T\rangle =
{\textstyle\frac{3}{4}}\,N_0 \,A(\lambda)\,R^{-2},\quad\hbox{where}\quad
A(\lambda)=(1+\lambda)({\textstyle\frac{5}{2}} +
\lambda)/\lambda;\label{eq:T}\end{equation}
\begin{equation}\langle U\rangle={\textstyle\frac{3}{4}}
\,N_0\,B(\lambda)\,R^2,\quad\hbox{where}\quad B(\lambda)
= ({\textstyle\frac{7}{2}}+\lambda)^{-1} \label{eq:U}\end{equation}
[note that the mean-square radius $\langle r^2\rangle$ is just  $2\langle
U\rangle/N_0 = \frac{3}{2} B(\lambda) \,R^2$,
which depends on both
$\lambda$ and $R$]; and
\begin{equation}\langle V\rangle = 2N_0\,{\cal
N}_0\,C(\lambda)\,R^{-3},\quad\hbox{where}\quad C(\lambda) =
\frac{\Gamma(3 +
2\lambda)}{\Gamma(\frac{3}{2})\,\Gamma(\frac{9}{2}+2\lambda)}\,
\bigg[\frac{\Gamma(\frac{7}{2}+\lambda)}{\Gamma(2+\lambda)}\bigg]^2.
\label{eq:V}\end{equation}
Thus, the variational energy functional becomes
\begin{equation}\frac{E_0(R,\lambda)}{N_0} =
{\textstyle\frac{3}{4}}\,A(\lambda)\,R^{-2}+ {\textstyle\frac{3}{4}}
\,B(\lambda)\,R^2+ {\cal N}_0\,C(\lambda)\,R^{-3}.\label{eq:E0var}\end{equation}
Since \(A\), \(B\), and \(C\) are all positive, this expression  obviously
has a minimum for some positive  \(R\), and the
derivative with respect to
\(R\) immediately gives the condition
\begin{equation}-A\,R + B\,R^5 = 2{\cal N}_0\,C.\label{eq:Rmin}\end{equation}
In addition, the derivative with respect to \(\lambda\) yields the second
condition
\begin{equation}{\textstyle\frac{3}{4}}\,A'(\lambda)\,R^{-2}+
{\textstyle\frac{3}{4}}
\,B'(\lambda)\,R^2+ {\cal N}_0\,C'(\lambda)\,R^{-3}=
0,\label{eq:lambdamin}\end{equation}
where \(C'(\lambda)/C(\lambda)\) is a linear combination of \(\psi\)
functions \cite{AS}.
Eliminating \({\cal N}_0/R^3\)
from Eqs.~(\ref{eq:Rmin}) and (\ref{eq:lambdamin}) leads to an {\it
explicit\/} equation for \(R\) as a function of
\(\lambda\):
\begin{equation}R^4 =
\frac{A}{B}\,\frac{C'/C-\frac{3}{2}\,A'/A}{C'/C+\frac{3}{2}\,B'/B},
\label{eq:R4}\end{equation}
and Eq.~(\ref{eq:Rmin}) then gives the corresponding value of \({\cal N}_0\).

\subsection{Large condensate (\({\cal N}_0 \gg 1\) and \(\lambda \ll 1\))}

As shown in Sec.~IV, these expressions are readily evaluated for any
positive \(\lambda\), but it is helpful first to
study two limiting cases, where analytic expressions can be obtained.  For
\(\lambda \ll 1\) and \({\cal N}_0 \gg 1\), a
detailed  expansion \cite{AS} yields
\begin{equation}C'/C\approx C_0  + C_1\,\lambda + \cdots,\end{equation}
where \(C_0 = \frac{3}{7} \) and \(C_1 = -\frac{2}{3}\pi^2 +
\frac{53}{9}+\frac{8}{25} +
\frac{16}{49}\approx -0.04432\), so that \(C'/C + \frac{3}{2}B'/B \approx
(C_1 + \frac{6}{49})\,\lambda + \cdots\,\).  It
is then easy to obtain the explicit result
\begin{equation}R^4\approx \frac{105}{8\,(C_1+
\frac{6}{49})\,\lambda^3},\quad\hbox{or}\quad R\approx
3.600\,\lambda^{-3/4},\label{eq:R}\end{equation}
along with \(n_0(r) = |\Psi_{TF}(r)|^2 \approx \frac{15}{8}\,(N_0/\pi
R^3)\,\big(1 - r^2/R^2\big)\) and the specific
results \(\mu_0\approx \frac{1}{2}\,R^2\), \(E_0/N_0 \approx \frac{5}{14}
\,R^2\),  and \({\cal N}_0\approx
\frac{1}{15}\,R^5\) given in the paragraph below Eq.~(\ref{eq:c0}).

\subsection{Small condensate (\({\cal N}_0 \ll 1\) and \(\lambda \gg 1\))}

In the opposite limit  \(\lambda \gg 1\) and \({\cal N}_0 \ll 1\), it is
simpler to return to Eq.~(\ref{eq:Rmin}), whose
solution is of the form
\(R({\cal N}_0) \approx R_0 + (C/2A)\,{\cal N}_0 + \cdots\,\) with \(R_0^4
= B/A\).  To first order in \({\cal N}_0
\ll 1\),   the condensate energy is simply
\begin{equation}\frac{E_0(\lambda)}{N_0} \equiv
\frac{E_0[R_0(\lambda),\lambda]}{N_0}=\frac{3}{2}
\,\sqrt{A(\lambda)B(\lambda)} +
\frac{C(\lambda){\cal N}_0}{R_0(\lambda)^3}.\end{equation}
 A direct expansion shows that \(A(\lambda)B(\lambda) \approx 1 +
\frac{5}{4}\lambda^{-2} + \cdots\,\), \(R_0(\lambda)
\approx
\sqrt{\lambda}\,(1  + \frac{7}{4}\lambda^{-1} +
\cdots\,)\), and
\(C(\lambda)  \approx
\lambda^{3/2}\,(2\pi)^{-1/2}\,(1+ \frac{69}{16}\lambda^{-1}+\cdots\,)\);
thus,
\begin{equation}\frac{E_0(\lambda)}{N_0} \approx \frac{3}{2} +
\frac{15}{8\,\lambda^2} + \cdots + \frac{{\cal
N}_0}{\sqrt{2\pi}}\, \bigg( 1 -
\frac{15}{16\,\lambda}+\cdots\,\bigg)\label{eq:gauss}\end{equation}
This expression clearly has a minimum as a function of \(\lambda \),
occurring at
\begin{equation}{\cal N}_0 \approx
\frac{\sqrt{32\pi}}{\lambda}\label{eq:smallN0};\end{equation}
the corresponding   condensate energy
\begin{equation}E_0({\cal N}_0 ) \approx N_0\,\bigg(\frac{3}{2} +
\frac{{\cal N}_0}{\sqrt{2\pi}}+
\cdots\,\bigg)\label{eq:smallE0}\end{equation}
precisely reproduces that found directly for the isotropic normalized
gaussian trial function \(\Psi_G(r) =
\sqrt{N_0}\,\pi^{-3/4}\exp(-\frac{1}{2}r^2)\) \cite{BP,cond-mat}.

\section{numerical results}
For any positive value of the variational parameter \(\lambda\), it is
straightforward to evaluate the other dimensionless
variational parameter  \(R\) from Eq.~(\ref{eq:R4}), the interaction
parameter \({\cal N}_0 = N_0 a/d_0\) that specifies
the condensate number from Eq.~(\ref{eq:Rmin}), and the various
contributions to the condensate energy per particle and
the condensate chemical potential. Typical results are shown in Table 1,
illustrating the continuous interpolation between
the TF  limit (large condensate with \({\cal N}_0 \gg 1\) and  \(\lambda
\ll 1\)) and the Gaussian limit (small condensate
with \({\cal N}_0\ll 1\) and
\(\lambda\gg 1\)).  In the TF limit, the small-\(\lambda\)  approximation
in Eq.~(\ref{eq:R}) yields
\(R\approx 113.8
\) for
\(\lambda = 0.01\), with the following TF values \({\cal N}_0 \approx
\frac{1}{15} R^5 \approx 1.275\times 10^9\) and
\(\mu_0 \approx \frac{1}{2} R^2 \approx 6480\);  these values are  close to
the respective  variational ones 114.4,
1.297\(\times 10^9\), and 6526 given in Table 1.  Similarly,   the Gaussian
(large-\(\lambda\)) limit takes \(R \approx
\sqrt\lambda\);  for  \(\lambda = 100\), this relation gives \(R = 10\),
and the approximate expressions in
Eqs.~(\ref{eq:smallN0}) and (\ref{eq:smallE0}) yield \({\cal N}_0\approx 0.100\)
    and
\(E_0/N_0 \approx 1.540\), which should be compared to the respective
variational values 0.105 and 1.541 from Table 1
(which also yields   the mean-square radius  $\approx 1.56$, close to that
for the ideal gas). Note that the
physical parameter
\({\cal N}_0\equiv N_0a/d_0\) is a single-valued function of
\(\lambda\) that decreases monotonically with increasing $\lambda$.

        As an example of the
utility of this variational approach, consider the situation for \(\lambda
= 1\), where   the ratio
\(\langle T\rangle/\langle E_0\rangle
\approx 0.022\) is already small (so that \(\mu_0 \approx \langle U +
V\rangle/N_0\)), yet the  variational condensate
density
\(|\Psi_{\lambda = 1}|^2 \propto (1-r^2/R^2)^2\)   differs greatly from the
parabolic TF condensate density
\(|\Psi_{TF}|^2\propto 1-r^2/R^2\). Thus the elementary criterion $\langle
T\rangle/\langle E_0\rangle\ll 1$ alone is
insufficient to ensure that the condensate density approaches the TF form.

 It is  important to recall that the variational method
provides only an upper bound for the total condensate energy, so that the
separate contributions are not necessarily
accurate.  In particular, Dalfovo, Pitaevskii, and Stringari
\cite{DPS} have used a boundary-layer formalism to show that the kinetic
energy per particle in the TF limit is
\(\propto R^{-2}\ln R\), which differs significantly from that found
variationally.

\section{behavior for Negative scattering length}

 The stability of a Bose condensate with  negative scattering length has
become of interest in connection with the
experimental study of \(^7\)Li \cite {Brad}. It is well known that a
uniform Bose condensate is unstable for arbitrarily
small negative interactions (the speed of sound becomes imaginary
\cite{Bog,FW});  this behavior is somewhat reminiscent
of Jeans's long-wavelength (gravitational)  instability for an infinite
uniform stationary medium \cite{Chand}, where  the
dispersion relation is simply that for a bulk plasma oscillation   with the
sign of the interaction reversed.  The
 external confining trap completely alters the situation, however, for it
eliminates   wavelengths longer than the
characteristic dimension of the trap (like electromagnetic standing waves
in a cavity).  Indeed, the Bogoliubov
dispersion relation  for a uniform medium with negative scattering length
$-|a|$ becomes \cite{Bog,FW} \(E_k^2 =
(\hbar^2
k/2m)^2\,(k^2  -16\pi |a|n_0)\), which is stable only for wavenumbers
greater than \(k_c=\sqrt{16\pi |a|n_0}\,\).  For an
order-of-magnitude estimate, take  uniform density and \(k_{\rm min}
\approx \pi/2d_0\) (appropriate for a spherical square well of radius
\(d_0\));  the condition $k_{\rm min} > k_c$
yields the approximate stability criterion
\(|{\cal N}_{0}| <{\cal N}_{0c}\equiv N_0|a|/d_0 = \pi^2/48\approx 0.206\).
For comparison, a numerical study of the GP
equation for negative scattering length in an isotropic  harmonic trap
\cite{RHBE} yields the   exact critical value
\({\cal N}_{0c} = 0.573\).

        In the present context, it is interesting to examine how the
variational calculation  treats a negative scattering
length.  Since the trap can remain stable at \(T\) = 0 K only for
relatively small condensates, it is appropriate to
consider the behavior for small negative values of \({\cal N}_0\).
Equation (\ref{eq:gauss}) remains valid, but the
altered sign of \({\cal N}_0\)
means that \(E_0(\lambda)\) achieves its minimum only for \(\lambda\to
\infty\), corresponding to a Gaussian trial
function.  This particular situation has already been examined in
\cite{cond-mat,Stoof}, where the local minimum in the
variational energy has been shown to disappear at the critical value
\(|{\cal N}_{0c}| =(8\pi/25\sqrt 5)^{1/2} \approx
0.671\), somewhat larger than the exact value found numerically in
\cite{RHBE}.

\section{Anisotropic harmonic trap}

The variational approach has the appealing feature that it is readily
generalized to an anisotropic harmonic trap
potential, with
\begin{equation}U({\bf r}) = \textstyle\frac{1}{2}
m\sum_j\,\omega_j^2r_j^2,\end{equation}
where $j = 1$, 2, and 3 runs over the three coordinate axes.
Correspondingly, the three (in general, distinct)
oscillator lengths are given by  $d_j^2 = \hbar/m\omega_j$.  As a natural
trial function, take
\begin{equation}\Psi({\bf r}) = c_0 \,\bigg(1 -
\sum_j\,\frac{r_j^2}{d_j^{\,2} R_j^2}\bigg)^{\!\! (1 +
\lambda)/2}\,\theta\bigg(1 - \sum_j\,\frac{r_j^2}{d_j^{\,2}
R_j^2}\bigg),\label{eq:aniso}\end{equation}
where \{$R_j$\} is a set of three dimensionless variational parameters.  The
normalization constant is easily evaluated with Eq.~(\ref{eq:norm}) by
going to ``scaled'' variables $x_j =
(r_j/d_j R_j)$, yielding the generalization of Eq.~(\ref{eq:c0})
\begin{equation}c_0^2 = \frac{N_0\,\Gamma(\frac{7}{2} +
\lambda)}{2\pi\,R_1R_2R_3
\,\Gamma(\frac{3}{2})\,\Gamma(2+
\lambda)}.\end{equation}

        The three relevant expectation values are also readily found for
this more general trial function
\begin{equation}\langle T\rangle =
{\frac{1}{4}}\,N_0
\,A(\lambda)\,\sum_j\,\frac{\hbar\omega_j}{R_j^{2}},\label{eq:Taniso}
\end{equation}
\begin{equation}\langle U\rangle={\frac{1}{4}}
\,N_0\,B(\lambda)\,\sum_j\,\hbar\omega_jR_j^2,\label{eq:Uaniso}
\end{equation} and
\begin{equation}\langle V\rangle =
2 N_0\,\hbar\overline\omega\,\frac{{\cal
N}_0\,C(\lambda)}{R_1R_2R_3},\label{eq:Vaniso}\end{equation} where
$\overline\omega^3 \equiv
\omega_1\omega_2\omega_3$, ${\cal N}_0 \equiv N_0a/(d_1d_2d_3)^{1/3}$, and
the functions $A$,
$B$, and
$C$ are the same  as for an isotropic trap, given in
Eqs.~(\ref{eq:T})--(\ref{eq:V}). Thus, the
general variational energy functional
\begin{equation}\frac{E_0({\bf R},\lambda)}{N_0} = {\frac{1}{4}}\,N_0
\,A(\lambda)\,\sum_j\,\frac{\hbar\omega_j}{R_j^{2}}+{\frac{1}{4}}
\,N_0\,B(\lambda)\,\sum_j\,\hbar\omega_jR_j^2
+ N_0\,\hbar\overline\omega\,\frac{{\cal
N}_0\,C(\lambda)}{R_1R_2R_3}\label{eq:E0varaniso}\end{equation}
 depends on the four variational parameters $R_1$, $R_2$, $R_3$, and $\lambda$.

It is not difficult to
 minimize this functional with respect to each of the first three
parameters. The three conditions $\partial
E_0/\partial R_j= 0$ yield  the equations
\begin{equation}-A/R_j^2 +BR_j^2  = 2D\overline
\omega/\omega_j,\label{eq:minRj}\end{equation}
 where $D \equiv {\cal
N}_0\,C(\lambda)/R_1R_2R_3$ is the same in each equation.  This set of
conditions  can be solved
formally to express  $R_j^2$ as
\begin{equation}R_j^2 =\frac{\overline\omega D}{\omega_j B}+
\sqrt{\frac{A}{B} + \left(\frac{\overline\omega D}{\omega_j
B}\right)^{\!\! 2}},\label{eq:Ej}
\end{equation}
where the sign of the root is chosen to give the correct limit for $a \to
0$ (the ideal gas).  Since $D^2 = ({\cal
N}_0C)^2/R_1^2R_2^2R_3^2$, these expressions  immediately provide one
relation between $D$ and $\lambda$.  In addition,  the minimum condition
$\partial E_0/\partial \lambda=0$
 can be combined with Eqs.\ (\ref{eq:minRj}) and (\ref{eq:Ej}) to provide a
second relation between $D$ and
$\lambda$, thus determining both quantities, as well as the three
dimensionless parameters $R_j $ and ${\cal
N}_0$.   Edwards {\it et al.\/}
\cite{EDCRB} and Dalfovo and Stringari
\cite{aniso} have used distinct  numerical procedures to evaluate the
ground-state condensate wave function for
anisotropic but axisymmetric traps, using the physical parameters
appropriate to \cite{And}; it would be interesting
to compare the present variational approach with their explicit results,
but such a numerical study remains to be
performed.

\acknowledgements
I am grateful to D. Rokhsar for valuable comments and suggestions.  This
work was supported in part by the National
Science Foundation, under Grant No.~DMR 94-21888.

\begin{table}
\caption{ Typical variational  physical parameters that characterize a spherical
condensate in an isotropic harmonic trap}
 \begin{tabular}{cccccccc}
\(\lambda\) &\(R\) & \({\cal N}_0\) & \(\langle T\rangle/N_0\) & $\langle
U\rangle/N_0 = \frac{1}{2}\langle
r^2\rangle$&${1\over 2}\langle V\rangle/N_0$ &$E_0/N_0$&$\mu_0$ \\ \hline
0.01&114.407&1.2974$\times10^9$&0.0145&2796.77&1864.51&4661.29&
6525.80\cr0.1&21.2428&2.685$\times10^5$&0.0475&94.0115&62.6426&156.702&219.3
44 \\
1&5.3967&153.92&0.1803&4.8541&3.1159&8.1503&11.2662 \\
10 & 4.3724 & 1.5756 & 0.5394 & 1.0621 & 0.3484 & 1.9499 & 2.2984 \\
100 & 10.3758 & 0.1054 & 0.7212 & 0.7801& 0.0393 & 1.5406 & 1.5799 \\
\end{tabular} \end{table}

\end{document}